\documentclass[preprint,tightenlines,superscriptaddress,
prd,nofootinbib,eqsecnum,showpacs]{revtex4}
 \usepackage{graphicx}
\def\beq{\begin{equation}} \def\eeq{\end{equation}}
\def\bea{\begin{eqnarray}} \def\eea{\end{eqnarray}}
\def\LP{L_{\rm P}} \def\lc{\ell_{\rm c}} \def\lq{\ell_{\rm q}}
\def\ld{\ell_{\rm d}} \def\ls{\ell_{\rm s}} \def\lg{\ell_{\rm g}}
\def\LM{L_{\rm m}} \def\man{{\cal M}} \def\ovr{\overline}
\def\dd{{\rm d}} \def\ee{{\rm e}}  \def\dontshow#1{}
\def\ev#1{\langle #1 \rangle}  
\def\lr{\leftrightarrow} \def\ts#1{{\textstyle{#1}}}\def\half{\ts{1\over2}}

\begin{document}

\title{Semiclassical Quantum Gravity:\\
Statistics of Combinatorial Riemannian Geometries}

\author{Luca Bombelli}\email{luca@phy.olemiss.edu}
\affiliation{Department of Physics and Astronomy\\
University of Mississippi, University, MS 38677, U.S.A.}
\affiliation{Perimeter Institute for Theoretical Physics\\
35 King Street North, Waterloo, Ontario, Canada N2J 2W9}

\author{Alejandro Corichi}\email{corichi@nuclecu.unam.mx}
\affiliation{Department of Physics and Astronomy\\
University of Mississippi, University, MS 38677, U.S.A.}
\affiliation{Department of Gravitation and Field Theory, Instituto
de Ciencias Nucleares,
Universidad Nacional Aut\'onoma de M\'exico,\\
A. Postal 70-543, M\'exico D.F. 04510, M\'exico\\{\quad}}

\author{Oliver Winkler}\email{owinkler@perimeterinstitute.ca}
\affiliation{Perimeter Institute for Theoretical Physics\\
35 King Street North, Waterloo, Ontario, Canada N2J 2W9}

\date{1 September 2004} 

\begin{abstract}
\noindent This paper is a contribution to the development of a framework,
to be used in the context of semiclassical canonical quantum gravity,
in which to frame questions about the correspondence between discrete
spacetime structures at ``quantum scales" and continuum, classical
geometries at large scales. Such a correspondence can be meaningfully
established when one has a ``semiclassical" state in the underlying
quantum gravity theory, and the uncertainties in the correspondence arise
both from quantum fluctuations in this state and from the kinematical
procedure of matching a smooth geometry to a discrete one. We focus
on the latter type of uncertainty, and suggest the use of statistical
geometry as a way to quantify it. With a cell complex as an example of
discrete structure, we discuss how to construct quantities that define
a smooth geometry, and how to estimate the associated uncertainties.
We also comment briefly on how to combine our results with uncertainties
in the underlying quantum state, and on their use when considering
phenomenological aspects of quantum gravity.
\end{abstract}
\pacs{04.60.Pp, 02.40.Sf, 05.90.+m.}
\maketitle

\section{Introduction}
\label{sec:1}

\noindent Ever since quantum mechanics and general relativity
became two of the pillars of contemporary physics, it has been
clear that a new, more general theory is needed that contains both
of them in some appropriate limit. The task of finding such a
quantum theory of gravity has indeed been an excruciating one and
has not yet been completed, in part because of the numerous
conceptual challenges that such a theory faces \cite{conceptual}.
One of them is the question that will motivate our considerations:
What is the fundamental structure of spacetime? That is, can we
expect the continuum picture of a (smooth) manifold to be valid at
arbitrarily small scales?  Do we need to modify it into a ``foamy"
picture with a manifold of fluctuating topology \cite{Wh}, and
possibly dimensionality? Or should it be entirely replaced by a
different, quantum picture that is discrete, polymeric, fuzzy?  Is
it even appropriate to pose the basic questions in such a
geometric language?  It is by now generally believed that the
smooth manifold picture is inadequate, but there is a wide variety
of approaches to the way in which the quantum realm manifests
itself, which in some cases depend on the prejudices of individual
quantum gravity practitioners, for example regarding the amount of
classical structures and non-dynamical elements that remain in the
final formulation.  In the present work we shall not pretend to be
free from such prejudices; however, we will attempt to attack our
main problem from a general perspective, and present a framework
that could be adapted to different approaches and formulations of
the dynamics.

Within the search for a consistent theory of quantum gravity,
there has been over the past several years an increase in the
amount of work devoted to the semiclassical sector of the theory
and its physical predictions. These subjects involve several
issues, such as identifying those states which have a
semiclassical interpretation, and predicting phenomenological
effects due to these states that can be seen as quantum gravity
corrections to the dynamics of quantum fields. As a result of
their different uses of the classical structures, each approach
faces a different challenge when it comes to the question of the
(semi-)classical limit. As loop quantum gravity \cite{lqg} is the
main inspiration for our framework, let us look at the
semiclassical limit there.

A key aspect, underlying the very possibility of phenomenological
predictions, is that of establishing a correspondence between
the structure in terms of which quantum gravity is formulated,
and the smooth geometry in terms of which the usual field
theories are formulated and observations interpreted.
One expects the two descriptions to be very close at large
scales, and to start departing significantly at small length
scales, since all of the field theories we use to describe the
behavior of matter, and even of gravitation, are based on
manifolds that become ``featureless" at small scales, and are
actually locally flat in an infinitesimal neighborhood of each
point.  We will therefore take the point of view that the manifold
$\man$ is only a convenient tool which gives us an effective
``low-energy" picture, within a certain degree of accuracy at each
scale. With this observation at the center of our discussion, we
will describe a framework that will quantify this accuracy, and
identify scales at which one can cross over between the quantum
picture of spacetime, assumed to be discrete, and a continuum,
classical geometry.

The issue of recovering a continuum theory as an appropriate limit
of a discrete one is of course not new, and has been studied in
various contexts for a long time. In quantum gravity the
best-known approaches in which the issue arises are the ones based
on piecewise linear manifolds, such as the various versions of
Regge calculus \cite{regge} or dynamical triangulations
\cite{loll}, graph-based approaches such as loop quantum gravity
\cite{lqg}, the more recently developed spin foam models
\cite{spinfoam}, or causal sets \cite{posets}, but the list
includes many others, both covariant and based on space+time
splittings. In the causal set approach and in dynamical
triangulations, the only variable used is combinatorial, while in
other approaches one uses a combinatorial structure which is
``dressed" with additional variables, such as edge lengths of a
manifold triangulation in Regge calculus, or holonomies of
connections along graph edges in loop quantum gravity. Although
there may or may not be a ``minimal length" in those frameworks,
one may think of their discreteness as giving rise to a
characteristic length scale $\ell$ (whose value in terms of
fundamental constants is to be determined within the theory) above
which the discrete structure $\Omega$ can be considered as
kinematically well approximated by a smooth manifold.

When one faces the issue of recovering the continuum theory,
the first question that arises is whether the discrete dynamics
converges to that of the continuum as $\ell \to 0$. To address
this question, one usually considers a fixed smooth manifold with
metric $(\man, g_{ab})$ (to be thought of as either the spatial or
the spacetime geometry), and embeds in it an unspecified, but
increasingly finer sequence of discrete structures $\Omega(\ell)$
(for example, triangulations of $\man$ with edge lengths induced
by $g_{ab}$ of order $\ell$), and shows that the value of the
discrete version of some quantity $Q$ that governs the dynamics,
such as the action in a covariant approach, approaches the value
of the continuum version, calculated for $(\man,g_{ab})$, as
$Q(\Omega(\ell)) = Q(\man,g_{ab}) + {\cal O}(\ell^2)$.  A number
of results of this type are known \cite{loll}.  For the most part,
the emphasis has been on features of the discrete dynamics itself
and on showing that the discrete variables approach the continuum
ones fast enough.

Here we want to explore the kinematical aspects of this
correspondence in more detail. As already mentioned, we feel that,
given the importance of potentially observable corrections to
continuum theories, it is important to have a way to quantify the
extent to which a discrete geometry corresponds to a smooth one,
and the amount of uncertainty in the correspondence. In other
words, if the discrete theory is fundamental and not an
approximation, the limit $\ell \to 0$ is {\it not\/} to be taken,
and we need to know exactly what the ${\cal O}(\ell^2)$ terms are.
This is the goal of the framework that we propose. In this work,
the first in a series of papers, we intend to motivate the issues
to be addressed and to provide the first steps. We start by
considering only those aspects that are related to an underlying
combinatorial structure; aspects related to additional variables
to be assigned to the underlying discrete structure will be
treated separately in a forthcoming publication \cite{bcw2}. All
considerations will be independent of the dynamics of the theory,
and in fact we don't need any details of the quantum theory,
except for occasional references to an underlying quantum state
$\Psi$, assumed to be semiclassical. This paper can be separated
in two parts. In the first one we develop a way of assigning a
smooth classical geometry to a given (random) graph. The second
part places this procedure in the more general context of building
a macroscopic geometry $(\man,g)$, starting from a semiclassical
state $\Psi_{\rm sc}$, or a graph $\Omega$ together with the
expectation values of observables ${\cal O}$, and raises the
related issues of coarse-graining and combining statistical
and quantum uncertainties.

Let us now comment on the way in which our work differs from
previous results on the continuum limit of a discrete theory. The
first key feature of our approach is related to the fact that,
rather than {\it reconstructing\/} a given smooth geometry, we are
interested in {\it constructing\/} one using a discrete structure
$\Omega$, which may not even be embeddable in a smooth geometry.
For example, in the picture of quantum geometry arising from loop
quantum gravity, the geometry at Planck length appears to be
distributional, with support on the edges of a graph. In the
current formulation of this approach, those graphs are embedded in
a given manifold $\man$, but the expectation is that it should be
possible to formulate it in terms of abstract graphs. The theory
should then include criteria for recognizing graphs which look
like manifolds at large scales, and specify how to determine the
emergent geometry and evaluate the uncertainties involved in the
construction. Several scales, associated with qualitatively
different descriptions of the geometry, will arise in this
process, and it should ultimately be possible to establish a
correspondence between those provided by continuum theories, such
as the Planck length $\LP$ at which classical geometry is expected
to break down, and the ones provided by the discrete theory, which
in the present paper are purely combinatorial and dimensionless,
such as the amount of coarse-graining necessary for $\Omega$ to be
embeddable.

There is a caveat, however: a random complex carries information
from the metric of the underlying manifold.  As will be shown in
detail, curvature quantities can be related to combinatorial
properties of the complex and vice versa. But this raises a
potentially worrisome point: would it be consistent to use a
complex that determines, in the above sense, a classical metric
$g_1$, say, but dress it with semiclassical states that are peaked
around a completely different, non gauge-equivalent classical
metric $g_2$? To circumvent this potential problem, it would be
convenient to have criteria to decide on the right random complex
to use for the semiclassical situation at hand. This will imply,
at this level, to use the quantum geometry $g_2$ implied by a
quantum state $\Psi$ and compare it with the geometry $g_1$
consistent with the discrete structure $\Omega$.

As tools for constructing a geometry $(\man, g_{ab})$, we will
identify examples of quantities $Q_X$ associated with appropriate
regions or submanifolds $X$ of $\man$; the second key feature of
our approach concerns the way in which uncertainties $\Delta Q_X$
are calculated using statistical techniques. Given one discrete
structure $\Omega$, the $\Delta Q_X$ represent the uncertainty in
the estimate of the effective geometry $(\man, g_{ab})$ that
$\Omega$ could be considered a discretization of. Such statistical
fluctuations can be, and have been, calculated in some cases,
using both analytical and numerical methods. Here we will use as
far as possible analytical techniques, in the spirit of the random
lattice approach to gauge theories in Minkowski space pioneered by
T D Lee and collaborators \cite{Lee}. Most of the explicit
calculations will be done in two dimensions; results in three or
more dimensions in general will have to be obtained numerically.

More specifically, in Sec \ref{sec:2} we define the setting of our
work, introduce and motivate the choice we make for $\Omega$, that
of a cell complex, and comment on the possibility of obtaining
cell complexes starting from graphs. Sec \ref{sec:3} contains the
main results in this paper; after an introduction, clarifying what
we mean by constructing an approximate geometry from $\Omega$, we
show explicitly how to carry out the construction in 2D. In Sec
\ref{sec:4} we return to more basic questions about our discrete
structure: we discuss the role of different length scales
in the discrete-continuum transition, and the more
general setting of structures that are not embeddable at a certain
length scale. For example, discrete structures associated with
candidate semiclassical states for quantum gravity will probably
not be generic graphs, and it seems reasonable to start exploring
the obstructions to graph embeddability by looking at cases that
are not too ``severe", where the discrete structure is, in an
appropriate sense, ``almost embeddable". To that end, we discuss
the notion of coarse-graining for a cell complex, intended to
produce in those cases an embeddable, smoothed-out version of
the discrete structure.

In Sec \ref{sec:5} we discuss the consequences of the quantum
fluctuations $(\Delta\hat Q_I)_\Psi$, and possible uses of our
results. As stated earlier, from the point of view of the
continuum, those fluctuations contribute to the total uncertainty
in the geometry $(\man, g_{ab})$; thus, on the one hand they will
ultimately allow us to quantify the goodness of $\Psi$ as a
semiclassical state, and on the other hand they will play an
important role in the relationship between $\Psi$ and continuum-based
phenomenology. Note that these modifications are of a different
nature than the corrections to field dispersion relations that
arise purely from discretization effects in Ref \cite{TH}. Finally,
we return to the motivation for this work, and discuss possible
applications in loop quantum gravity phenomenology and other
directions for future work.

Regarding notation, we will follow the following convention.
Statistical averages and quantum expectation values will be
indicated by angle brackets, as in $\ev Q$, and means with respect
to probability distributions by overbars, as in $\ovr Q$; as for
uncertainties, $(\Delta Q)_\Psi^2$ will denote a quantum
fluctuation, while $(\Delta Q)^2$ (the subscript ``c" being
understood) or $\sigma^2_Q$ will denote the statistical
uncertainty or variance of a classical probability distribution.

\section{Mathematical Setting}
\label{sec:2}

\noindent Our framework can be seen as a ``bridge" between a
discrete, pre-geometrical description of spacetime, motivated
by quantum gravity ideas, and the classical, continuum-based
geometrical description, for situations in which the quantum
theory provides us with a ``semiclassical state". In this section,
we define the notions needed to translate this statement and
the conceptual points discussed in the introduction into a
specific program.

\subsection{Cell Complexes and Tilings}
\label{sec:2.1}

\noindent The most basic variable in this paper will be a cell
complex; given their central role in what follows, we start by
recalling a few useful definitions and facts about cell complexes
and their relationship with manifolds. In topology, a $k$-dimensional
(open) {\it cell\/} is a space homeo\-morphic to the interior of a
$k$-ball. A {\it cell complex\/} is a set of nonempty, pairwise
disjoint cells, such that (a) The closure of each cell is homeomorphic
to a ball and its boundary to a sphere in some dimension, and (b) The
boundary of each cell is a union of cells; in our case, this will
always be a finite union. (0-dimensional balls are pairs of vertices.)

Given a differentiable manifold $\man$, a cell decomposition
of $\man$ is a cell complex homeo\-morphically embedded in it.
Our assumptions then imply that the cell decomposition is locally
finite, in the sense that every compact subset of $\man$
intersects only a finite number of cells. For example, a finite
3-dimensional cell complex $\Omega$ (one whose maximal cell
dimensionality is 3) consists of a set of $N_0$ vertices $v_I$,
$N_1$ edges $e_I$, $N_2$ 2-cells $\omega_I$, and $N_3$ 3-cells $C_I$;
if $\Omega$ is a cell decomposition of $\man$, then the 3-cells
together with their boundaries, $(\bigcup_I C_I) \cup (\bigcup_I
\omega_I) \cup (\bigcup_I e_I) \cup (\bigcup_I v_I) \simeq \man$.
We will say that a cell complex $\Omega$ is {\it embeddable\/} if
there is a differentiable manifold $\man$ of which $\Omega$ is a
cell decomposition.

We will also mention the more general concept of tiling of $\man$.
This term often denotes a collection of pairwise disjoint open
subsets $\omega_I$ of $\man$ whose closures cover $\man$; here we
will consider a tiling to include the union of the boundaries of
the $\omega_I$, partitioned into submanifolds of various dimensionalities.
A simple example will illustrate the concept. Given a smooth loop
$\alpha$ in S$^2$, such that S$^2 \setminus \alpha$ consists of two
open ``half-spheres" $\omega_1$ and $\omega_2$, the set $\{\omega_1,
\omega_2, \alpha\}$ is a tiling of S$^2$, but not a cell decomposition.
However, if we pick a point $v \in \alpha$ and call $e$ the edge
$\alpha \setminus v$, then ${\cal C}_1:= \{ \omega_1, \omega_2,
e, v \}$ is a cell decomposition of S$^2$; alternatively, if we
pick two points on $\alpha$ and join them with an extra edge that
does not meet $\alpha$ elsewhere, we divide S$^2$ into three
open wedges which, together with the elements on their boundaries,
make up another cell decomposition ${\cal C}_2$. In a cell complex
$\Omega$ homeomorphic to a $D$-manifold $\man$ (but not in any tiling),
the $N_k(\Omega)$ satisfy
\beq
      \sum\nolimits_{k=0}^D (-1)^kN_k(\Omega)
      = (-1)^D\,\chi(\Omega)\;,\label{euler}
\eeq
where $\chi(\Omega)$ is the Euler number of the complex $\Omega$,
or of the manifold $\man$.

However, cell complexes need not be embeddable. They could,
for example, have ``regions" with different dimensionalities; in
an embeddable cell complex every cell is, or is on the boundary
of, one of maximal dimensionality. For a different type of example,
consider the above cell complexes ${\cal C}_1$ and ${\cal C}_2$,
and remove their 2-dimensional cells; in the first case the
remaining 1-dimensional skeleton is homeomorphic to S$^1$, while
in the second case one is left with a non-embeddable complex.

A useful operation on cell decompositions is duality, which
produces a new $D$-dimensional cell complex $\Omega^*$ from
any given $\Omega$ of the same dimensionality. One associates
with each $k$-dimensional cell $\omega$ in $\Omega$ (a
$D$-dimensional cell complex must have cells of all dimensionalities
$0 \le k \le D$) a $(D-k)$-dimensional dual cell $\omega^*$ in
$\Omega^*$, whose boundary consists of the duals of all cells
which have $\omega$ on their boundary. If $\Omega$ is a cell
decomposition of a manifold $\man$, then $\Omega^*$ is also
homeomorphic to $\man$; however, since the duality $\Omega
\leftrightarrow \Omega^*$ is an operation between abstract cell
complexes, in general there is no natural embedding of $\Omega^*$
in $\man$ (the mapping $f: \Omega \to \man$ does not induce a
mapping $f^*: \Omega^* \to \man$, unless $\man$ is endowed with
more structure). Duality is defined for general tilings, but
their duals may not be homeomorphic to the original $\man$, while
non-embeddable cell complexes may not have well-defined duals.

\subsection{Triangulations and the Voronoi Procedure}
\label{sec:2.2}

\noindent Of all types of cell decompositions of manifolds,
the most useful ones for us are triangulations, in which the
cell complexes are simplicial, and their dual complexes. In
triangulations, of course, all 2-cells (triangles) have 3 edges
and 3 vertices, all 3-cells (tetrahedra) have 4 faces, 6 edges,
and 4 vertices; in general, all $k$-simplices have $k+1$ faces
on their boundary, etc. Their dual cell decompositions therefore
satisfy incidence properties which state that each $(D-k)$-cell
is on the boundary of (is shared by) $k+1$ cells of dimensionality
one unit higher, etc, and can be concisely written as follows:
For each $l$-dimensional cell $\omega \in \Omega$, the number of
$k$-cells that have $\omega$ in their closure (with $0 \le l \le
k \le D$) is
\beq
    N_{k|l}(\omega) = {D+1-l\choose k-l}\;. \label{sharing}
\eeq
In particular, each vertex has $N_{1|0} = D+1$ edges. Thus, in
two dimensions all dual vertices are trivalent, while in three
dimensions they are shared by four edges. This property already
makes such complexes useful, since for example quantum geometry
results in loop quantum gravity show that 4-valent vertices of
graphs in 3 dimensions are the fundamental units of volume
\cite{vol}.  Also, if each edge terminates at two vertices,
we find a useful relation between the total numbers of edges
and vertices (if finite),
\beq
      N_1(\Omega) = \half\,(D+1)\,N_0(\Omega)\;. \label{edges}
\eeq
However, the main reason why these two complexes are useful
is more general; it lies in the fact that they can be obtained
in a manifold using just a set of points as input, and in the
way they encode geometrical information on $(\man, q_{ab})$
when the set of points they are based on is chosen at random.

Given any locally finite set of points $p_I$ in a Riemannian
manifold $(\man, q_{ab})$, one can obtain from it a tiling
of $\man$; if the points are at generic locations and
sufficiently dense (with respect to both local length scales,
determined by the metric, or global ones, determined by the
metric and topology), the result is actually
a cell decomposition, called the Voronoi complex, and its dual
is called the Delaunay triangulation. We start by introducing
the procedure in general, without any additional assumptions on
the set of points, and then consider in the next subsection the
case in which the $\{p_I\}$ are randomly sprinkled points.
(Capital latin indices $I$, $J$, ..., will be used to denote
points or elements of various dimensionalities in a complex;
the type of object they refer to should be clear from the context.)

For each embedded point $p_I$, we can define an open region
$\omega_I \subset \man$ as the set of all manifold points which
are closer to $p_I$, with respect to $q_{ab}$ than to any other
$p_J$; clearly the union of the closures of such regions is
$\man$, so the $\{\omega_I\}$ define a tiling.
The boundaries of the $\omega_I$ are made of manifold points that
are equidistant from more than one of the $p_I$'s; those equidistant
from $p_I$ and $p_J$ and closer to them than to any other $p_K$ are
the codimension-1 common boundary of the $D$-dimensional regions
$\omega_I$ and $\omega_J$ around $p_I$ and $p_J$, respectively.
Common portions of boundaries among more than two cells, if present,
define lower-dimensional portions of the $\partial\omega_I$'s.
We will always call the resulting complex a {\it Voronoi complex,}
even when not all the sets just described are cells; if the $p_I$
are close enough to each other compared to all length scales in
the manifold, associated with the metric or the topology, we
actually obtain a cell complex $\Omega$ homeomorphic to $\man$.

\begin{figure}
  \includegraphics[angle=0,scale=.9]{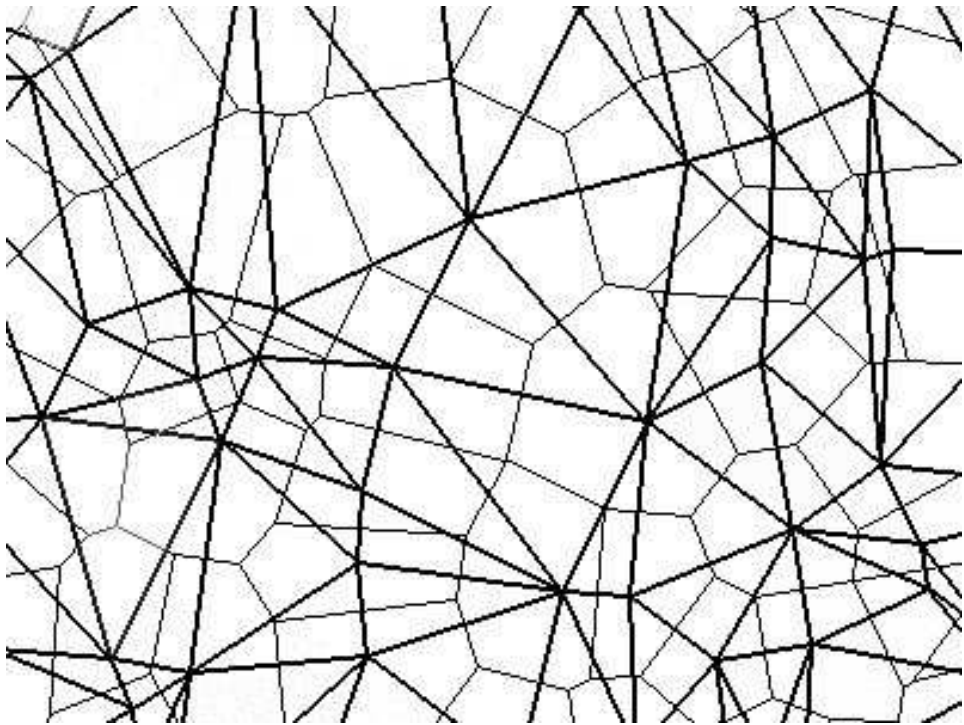}
  \caption{\label{f1}
  Example of 2D Voronoi (thin lines) and Delaunay (thick lines)
  complexes; the points they are based on are the Delaunay vertices.
  Notice that in 2D edges are dual to edges, whereas vertices
  and 2-cells are duals of each other, and that dual edges are
  orthogonal to each other, although they do not necessarily meet.}
\end{figure}

Generically, vertices of a Voronoi complex are equidistant
from $D+1$ points, since a higher number of points at the same
distance can only be obtained in degenerate situations; then the
$D+1$ ways of picking $D$ of those points define $D+1$ Voronoi
edges incident on that point. In fact, when the $p_I$ are at
generic locations, all of the relationships (\ref{sharing}) are
satisfied, regardless of whether we have a cell complex; because
in our applications the points will be randomly sprinkled in $\man$,
we will not discuss special arrangements and will always assume
that those relations hold.
Given any Voronoi vertex $v_I$, the $D+1$ points it is close to are
on the surface of a $D$-sphere around $v_I$ and define a simplex
$\omega^*_I$, dual to $v_I$. If $\Omega$ is a cell complex, there
are enough Voronoi vertices for the dual complex $\Omega^*$ defined
by all those simplices to be a {\it Delaunay triangulation\/}
of the manifold, which has the $p_I$ themselves as vertices.
Intuitively, $k+1$ sprinkled points that are ``clustered closely
enough" define a $k$-simplex in $\Omega^*$ that lies in the
$k$-plane through those points. Thus, the concept of Delaunay
triangulation is less general than that of ``Voronoi tiling".

When does the Voronoi procedure not produce a cell complex?
A simple example is that of a 2-sphere with two points $p_1$ and
$p_2$ chosen on it, in which case the procedure gives the tiling
we called $\{\omega_1, \omega_2, \alpha\}$ in Sec \ref{sec:2.1},
and its dual consists of $p_1$, $p_2$ and an edge between them,
which is not homeomorphic to S$^2$. In this example, both the
non-trivial topology and the high curvature (in terms of the point
density) contribute to the outcome. However, one can easily modify
it into other examples in which only one of the factors is present
(a flat cylinder, or a plane with a small region blown up into a
long tube or a balloon, respectively), and it produces a similar
effect. All such Voronoi tilings satisfy the incidence relations
(\ref{sharing}), and those same relations, imposed on an abstract
complex, guarantee that it is homeomorphic to some manifold.
For the time being, however, we would like to work with ordinary
cell complexes. This means that for us an embeddable complex will
be one such that the relations (\ref{sharing}) are satisfied and
in which every cell has on its boundary (the appropriate number
of) cells of all lower dimensionalities.

By construction, the two types of complexes just defined are
embeddable. However, since in this paper our intention is to
{\sl construct\/} geometries from discrete structures, we are
interested in characterizing those structures which can arise
from the Voronoi procedure in some Riemannian manifold. This
condition excludes the ``lower-dimensional parts" in the complex,
but other possible ``defects" are not excluded. Voronoi complexes
provide the most convenient combinatorial set of conditions for
being embeddable in a manifold. These conditions exclude the
``degenerate" Voronoi complexes mentioned earlier (which is not
a big loss), but they do include ones which are not locally
finite,  and could easily be extended to the generalized
complexes that are not made of topological cells.

\subsection{Random Voronoi Complexes}
\label{sec:2.3}

\noindent In order to understand how discrete structures
embedded in a manifold $\man$ encode its geometry, we need
to introduce a few notions related to randomly distributed
points. A random point distribution on $\man$ with a volume
element (in particular, one given by a Riemannian metric,
$\sqrt g$) is the outcome of a uniform, binomial or Poisson,
point process. If the total volume $V_\man$ is finite, a
uniform point process is specified by stating that, each
time a point $\bf x$ is chosen in $\man$, the probability
that $\bf x$ fall in any given measurable region $X\subseteq
\man$ is
\beq
    P({\bf x}\in X) = V_X/V_\man\;, \label{PxinX}
\eeq
or the infinitesimal version, that the probability density is
$\tilde P_\man({\bf x}|\sqrt g) = V_\man^{-1} \sqrt g\,\dd^D x$.
If the process is repeated $N$ times, with no correlations among
points (we will not keep track of the order they came in), we get
a uniform sprinkling of points with density $\rho:= N/V_\man$
that for our purposes we will think of as being of order $\lc^{-D}$.
The probability density for each point to fall in an infinitesimal
region $\dd^Dx$ is
\beq
    \dd\mu = \rho\,\sqrt{g}\,\dd^Dx\;.
\eeq
One of the most useful finite probabilities in this context is the
one for exactly $k$ points out of $N$ to fall inside $X$ (without
specifying which ones). It is easy to see that this probability
follows a binomial distribution,
\beq
    P(k,X\mid N,\man)
    = {N\choose k} \left({V_X\over V_\man}\right)^{\!k}
    \left(1-{V_X\over V_\man}\right)^{\!N-k}, \label{PkinX}
\eeq
which, as $V_\man$ and $N$ become very large, with $\rho =$ constant,
approaches a Poisson distribution,
$$
    P(k,X\mid N,\man) \approx {\ee^{-\rho V_X}\,(\rho\,V_X)^k\over k!}\;.
    \label{Poisson}
$$
This last equation justifies the name {\it Poisson distribution\/}
that is often used for the sets of points used in this paper, and
corresponds to the infinite volume situation.

A random Voronoi complex in a manifold $(\man, g_{ab})$ is the
result (assumed to be a cell complex for the time being) of
applying the Voronoi procedure of Sec \ref{sec:2.3} to a Poisson
distribution of points in $(\man, g_{ab})$ (see, e.g., Ref \cite{Sa}
and references therein). This is the discrete structure we use in
this paper to encode information on a spatial geometry. Such
structures have been called ``random lattices" in the context
of gauge theory (the subject is actually older than that, but
for references with physical motivations somewhat related to
ours, see Refs \cite{Lee} and \cite{ID}). The randomness of
the point distributions and their finite density imply that all
complexes are locally finite, and all vertices are $(D+1)$-valent
with probability 1. Most of the general discussion will be valid
for complexes and manifolds of arbitrary dimension $D$, but
actual calculations will be carried out for 2-dimensional ones.

\subsection{Remarks on the Use of Voronoi Complexes}
\label{sec:2.4}

\noindent The discrete structures one uses in loop quantum gravity
to construct spin networks, the basic states that form the usual
basis for the kinematical Hilbert space ${\cal H}_{\rm kin}$, are
graphs embedded in a manifold. Thus, although we don't know yet
how to formulate a manifold-independent theory, it may be useful
to try to establish a correspondence between certain types of
graphs and the Voronoi cell complexes that our statistical
machinery is based on. Thus, suppose that we are given a graph
$\gamma$, consisting only of vertices and edges, but without
higher-order cells. We then pose the following question: Can we construct
a full Voronoi complex just from $\gamma$? To recover $\Omega$ as
an abstract complex in $D$ dimensions, what we need to do is specify
which edges in $\gamma$ form (the boundary of) an elementary 2-cell,
which of the latter form (the boundary of) a 3-cell, and so on. If
all goes well, the resulting cell complex will satisfy the correct
incidence relations for a Voronoi complex, as specified by
(\ref{sharing}).

Let us begin with a proposal for a construction in two dimensions.
Given a graph $\gamma$ in which every vertex is trivalent, define
a {\it loop\/} to be a chain of consecutive edges $e_1e_2 \cdots
e_K = (v_1\lr v_2\lr \cdots \lr v_K\lr v_{K+1})$ that closes on
itself, i.e., $v_1 = v_{K+1}$.  Most loops are not to be thought
of as boundaries of 2-cells; we call {\it plaquette\/} a loop $\alpha$
such that, for any two vertices $v_I$ and $v_J\in\alpha$, the
shortest path in the graph between $v_I$ and $v_J$ is part of
$\alpha$. The cell complex $\Omega$ we are looking for has as 0-cells
and 1-cells the same ones as $\gamma$, trivially, and its 2-cell are
identified with a set of plaquettes $\alpha$ such that $\Omega$ is
a Voronoi cell decomposition of a 2-manifold, i.e., such that every
edge is shared by exactly two 2-cells; if such a choice is not possible,
we consider the graph non-embeddable. However, we expect that if
we apply this construction to the 1-skeleton of an actual Voronoi
complex $\Omega$, we recover the original $\Omega$.
The main questions will then be: How do we recognize from the
graph whether it corresponds to a ``good situation"? What can
we say about cases where it does not?

Similarly, to construct a 3D Voronoi complex from a four-valent
graph, we start by defining a candidate 3-cell $C$ as a finite set
of plaquettes $\{\alpha_1,\alpha_2,\dots,\alpha_m\}$ such that
(i) every edge is shared by exactly two 2-cells $\alpha_i$ and
$\alpha_j$, as in the 2D construction above, (ii) the topology
of $C$ is that of a 2-sphere, and (iii) for every two vertices
$v_i \in \alpha_i$ and $v_j \in \alpha_j$, the shortest path in
the graph between them is part of $C$. A collection of 3-cells
defined in this way gives a good Voronoi complex if all edges in
it are shared by exactly three plaquettes, and each plaquette by
exactly two 3-cells. These definitions could then be generalized
in an obvious way to higher dimensions, but in this paper we will
not need to consider explicitly cells of higher dimensionality.
We can call {\it abstract $D$-dimensional Voronoi graph\/} one
such that all vertices have the same valence $D+1$, and the
above construction gives a good $D$-dimensional Voronoi complex.

Let us conclude with two remarks on the concepts we have
introduced so far. First, we are not suggesting that all
semiclassical quantum gravity states are associated with Voronoi
complexes, just as in ordinary quantum mechanics, not all
semiclassical states are coherent states. However, the latter have
properties that make them easy to work with, and they encode in a
convenient, minimal set of parameters a point in classical phase
space and the freedom in the (minimum) uncertainties in the
canonical variables. We propose to consider states based on
Voronoi complexes as playing a similar role for the
discrete-to-continuum transition. We do not know yet how to phrase
a minimum uncertainty condition in this context, or questions
about the existence of processes which might produce such states.
Filling in the first gap is one of the goals of our program; the
second one will probably require a much more complete knowledge of
semiclassical quantum gravity, including its dynamics. From a
geometrical point of view, however, there is a strong motivation
for using Voronoi complexes, that we will be exploring in this
paper.

Second, it is known that in order for states based on
a set of graphs to span a dense subset of the
(kinematical) Hilbert space of loop quantum gravity,
one needs to consider graphs with an arbitrary number of edges
and connectivities. If one restricts oneself to graphs with only
four-valent vertices (in 3+1 dimensions), one does not obtain
but a high-codimension subspace of the Hilbert space. We are
nevertheless suggesting that by restricting our attention to
states defined over such graphs we will not lose important
information. Are the semiclassical states defined over such a
restricted class of graphs sufficient to display the needed
semi-classical features? We do not have a definite proof for this,
but we can argue in favor of such states. Consider for instance
the example of a simple harmonic oscillator with a finite number
of degrees of freedom. In this case the usual Gaussian coherent
states one defines, peaked at phase space points, span only a
finite-dimensional submanifold of the infinite-dimensional Hilbert
Space of the theory. In spite of this, one can regard the coherent
states as `enough' for describing semiclassical states in some
cases. Again for the coherent states of the free Maxwell theory,
one can take coherent states and they approximate very well the
semiclassical properties that we are interested in. We will then
by analogy assume that the states we are considering here, defined
over Voronoi complexes, will be enough to describe the
semiclassical sector of the theory.

\section{The Intrinsic Geometry of Voronoi Complexes}
\label{sec:3}

\noindent Having introduced the necessary background concepts,
we can now describe in more detail what we intend to do in the
rest of the paper. Our general goal is the following: given a
Voronoi cell complex $\Omega$, i.e., one satisfying the incidence
relations (\ref{sharing}), determine the range of classical
geometries that are consistent with it. From the quantum gravity
point of view, it is not clear whether it is reasonable to associate
a single $\Omega$ with a semiclassical $\Psi$ (different points
of view underly for example the proposal in Ref \cite{Bo}, or the
shadow state proposal \cite{shadow}, in which different discrete
structures are seen as tools for probing the state $\Psi$).
We do so here because it allows us to separate the effects of
the classical uncertainty in the discrete-to-continuum transition
from those of the quantum fluctuations in $\Psi$, and if more
$\Omega$'s need to be considered one can always combine their
uncertainties later.

We emphasize that the effective geometry will only be a spatial
one; the recovery of an effective spacetime geometry requires
either structures that can be interpreted as discrete spacetimes
and the use of Lorentzian statistical geometry (see, e.g., Ref
\cite{JMP}), or additional variables on $\Omega$ that can be
interpreted as discrete versions of dynamical data (as in Ref
\cite{bcw2}). Even in this context, our actual calculations will
concern cases in which the topology of $\man$ is trivially determined
by that of $\Omega$, and only Sec \ref{sec:4} will discuss a more
general situation.

\subsection{The Discrete-Continuum Transition}
\label{sec:3.0}

\noindent There is an analogy with our situation in elementary
physics. When one is dealing with a fluid, one `knows' that at some
`microscopic scale', one is dealing with molecules, individual
entities with which one can associate, classically, a position
and a velocity. One then considers cells at a mesoscopic (crossover)
scale inside of which one averages velocities, energies and so on,
and one assigns such quantities to the cell as a whole. Finally,
one goes to much larger, macroscopic scales and regards those
properties of the cell as being local, defined by continuum (and
differentiable) fields. In a sense, this is the procedure we
are envisaging: There is a microscopic (discrete scale) $\ld$
where the `true' discrete geometry is defined by a graph. We will
then consider large sets of cells at a scale $\lc$, over which we
will average the combinatorial quantities of the cells, to smooth
out statistical fluctuations and define mesoscopic quantities
that vary slowly between such groups of cells. On the larger
macroscopic scale, this will allow us to view the mesoscopic
quantities as the local values of continuum fields that define
a geometry. This passage between the microscopic scale $\ld$
and the final macroscopic one is what we call the
{\it discrete-continuum\/} transition.

To define a macroscopic geometry $(\man,q_{ab})$, the quantities
we consider will be geometric invariants $Q_X$ associated with
extended submanifolds $X \subset \man$, large enough to correspond
to a large number of cells of $\Omega$ considered as a cell
decomposition of $\man$, but macroscopically small so that they
do not correspond to integrating or averaging over regions where
the geometry varies. In this paper, the submanifolds $X$ will be
simply open regions of $\man$ containing large numbers of $D$-cells
in $D$ dimensions, although for other purposes \cite{bcw2} one might
consider hypersurfaces in $\man$, approximated by large collections
of $(D-1)$-cells, or submanifolds of higher codimension.  As for
the $Q_X$ themselves, the above analogy with the thermodynamic
limit leads us to divide the possible invariants into extensive
ones (the simplest example is the volume $V_X$, for which we obtain
values by counting either Voronoi cells or Voronoi vertices
contained in $X$) and intensive ones (examples of this type are
curvature invariants, averaged over $X$). For the purposes of this
paper, the latter are the less trivial and the more fundamental
ones (relationships between most quantities are curvature-dependent),
and we will concentrate on those in this paper. By constructing
an effective continuum geometry here we thus mean finding the
values of a sufficiently large set of curvature invariants $Q_X$,
and providing a quantitative measure of the goodness of the
construction, in terms of the values of the uncertainties
$\Delta Q_X$, using $\Omega$.

In order to learn how to do this, we will first provisionally
assume that a classical geometry with known values for all of its
$Q_X$ is given, which allows us to calculate statistical distributions
of Voronoi complex variables. The relationship will then be inverted
to allow us to estimate the $Q_X$ and their uncertainties from a
given $\Omega$. In other words the main idea is that, when one obtains
a cell complex as the result of applying the Voronoi construction to
a random point sprinkling of density $\rho$ in a Riemannian manifold
$(\man, q_{ab})$, the combinatorial quantities $N_{k|l}$ for the complex
satisfy dimension-dependent identities, and one can also calculate
(at least in principle) geometry-dependent probability distributions
for values of those quantities. These two types of relationships
together imply that the complex encodes enough information about
the manifold $\man$ and metric $q_{ab}$ that we could reconstruct
$(\man, q_{ab})$ from it, up to statistical uncertainties on
volume scales at or below $\rho^{-1}$.

Thus, if we are given an abstract, embeddable Voronoi complex
in the sense of Sec \ref{sec:2}, we can {\it construct\/} an
approximate $(\man, q_{ab})$ that is a good continuum version
of $\Omega$ on scales larger than the average embedded cell size
$V_\man/N_D =: \rho^{-1}$. From a practical point of view, we quickly
run into the difficulty that only a few of the relevant probability
distributions, for low-dimensional flat or constant-curvature spaces,
are known. Therefore, we will treat in detail the two-dimensional
case, where we can derive the results we need analytically, and
outline the procedure in three dimensions, where analogous
calculations will have to be done with computer simulations.

\subsection{The Two-Dimensional Case}
\label{sec:3.1}

\noindent In two dimensions, any metric is conformally flat, and
can be locally written as $q_{ab} = \ee^{2f} e_{ab}$, where $f$
is a scalar function, and $e_{ab}$ a fixed flat metric on a portion
of the 2-manifold; if Cartesian coordinates are used for $e_{ab}$,
the line element is then $q_{ab}\,\dd x^a\dd x^b = \ee^{2f}\,
(\dd x^2 + \dd y^2)$. The function $f$ is in turn related to the
scalar curvature by $R = -2\,\ee^{-f}\nabla^2f$; $R$ is the continuum
geometrical quantity we will associate with a cell complex in this
section. When attempting to construct a geometry, it would seem
natural to consider first finding a distance for any pair of cells
(or of vertices) in the complex. However, the relationship between
areas and lengths is curvature-dependent, and if we consider the
cell density to be a basic parameter, before we assign distances
to pairs of objects in the cell complex, we first need to find out
the curvature that best fits each portion of $\Omega$. Besides,
in physical applications one is often directly interested in
the curvature of a manifold, since it affects, e.g., the
propagation of matter fields on it.

In order to find out how to assign a value of $R$ to a subset
of a cell complex $\Omega$, we need to learn to recognize
complexes that might arise from a point sprinkling in a manifold
with curvature $R$. We will therefore  use Eqs (\ref{euler}) and
(\ref{edges}) to determine the mean and variance of the number
of edges of a 2-cell on a 2-sphere $({\rm S}^2, s_{ab})$ of
constant scalar curvature. If we denote by $\Omega_R$ a Voronoi
complex on such an S$^2$, Eq (\ref{euler}) becomes $N_0-N_1+N_2
= \chi(\Omega_R)$; then, using (\ref{edges}) and noticing that
the average value of $N_1(\omega)$ over the complex is given by
$\ev{ N_1(\omega)}_{\omega \in \Omega_R} = 2\,N_1(\Omega_R)/
N_2(\Omega_R)$, where the 2 is due to the fact that each edge
is shared by two 2-cells, we obtain
\beq
    \ev{N_1(\omega)}_{\omega\in\Omega_R}
    = 6\left(1-{\chi(\Omega_R)\over\rho V}\right)
    = 6\left(1-{\int_\man R\,\dd v\over4\pi\rho V}\right)
    = 6\left(1-{R\over4\pi\rho}\right). \label{2D1}
\eeq
Here, we have used the Gauss-Bonnet theorem relating
$\chi(\Omega_R)$ and the scalar curvature of $({\rm S}^2, s_{ab})$.
Since the average does not depend on $\Omega_R$, it also equals
the mean number $\ovr{N_1}$ of neighbors of a cell taken over
all random complexes in the geometry $({\rm S}^2, s_{ab})$ with
density $\rho$. Eq (\ref{2D1}) can then be easily inverted to give
the scalar curvature in terms of the mean number of edges of a cell,
\beq
    R = 4\pi\rho\, (1-\textstyle{1\over6}\,\ovr{N_1})\;. \label{Rest}
\eeq

Finding the variance $\sigma_1^2 = \ovr{N_1^2} - (\ovr N_1)^2$ of the
distribution of the number of neighbors is considerably more difficult.
We already know $(\ovr N_1)^2$; for $\ovr{N_1^2}$, we will use a trick
\cite{Br}. Notice that for any 2-cell $\omega$, $N_1(\omega)$ equals the
number of vertices $N_0(\omega)$, so our task can be seen as that of
calculating $\ovr{N_0^2}$. The latter has been calculated (i.e., analytically
reduced to an integral which is then numerically evaluated) by Brakke
\cite{Br} for the flat case; we will now generalize his calculation to
the case of a constant positive curvature manifold. For any $\omega$,
a quantity related to $N_0^2$ whose mean is much simpler to calculate
directly is the number of (unordered) pairs of vertices not sharing an
edge, for if we denote this quantity by $N_{0,0'}(\omega)$, then by
definition $N_{0,0'} = \half\, N_0\, (N_0-3)$. Thus, $\ovr{N_0^2} =
2\,\ovr{N_{0,0'}} + 3\, \ovr{N_0}$ (notice that our $\ovr{N_{0,0'}}$ is
Brakke's $I(v,v)$), where $\ovr{N_{0,0'}}$ can be found by integrating
a suitable probability density, as we now show.

Consider the 2-sphere as embedded in 3-dimensional Euclidean space,
and call $C$ the center of the sphere; the radius $a$ of the sphere
is related to the scalar curvature of S$^2$ by $R = 2/a^2$. Choose an
arbitrary point $O$ on the 2-sphere as its origin. To locate any other
point $P$ in S$^2$ we will initially use the spherical coordinates
$(\chi,\theta)$, where $\chi \in [0,\pi]$ is the angle at $C$ between
the lines $CO$ and $CP$, and $\theta \in [0,2\pi]$ the azimuthal angle
on S$^2$ around $O$. The line element on S$^2$ is then given by the
familiar form
\beq
    \dd s^2 = s_{ab}\, \dd x^a\dd x^b
    = a^2\,(\dd\chi^2+\sin^2\chi\,\dd\theta^2)\;.
\eeq
Consider now an arbitrary cell in a random Voronoi complex on
(S$^2, s_{ab}$) of density $\rho$, and for convenience choose the
coordinates such that the sprinkled point or ``seed" $S_0$ that
defines this cell is at the origin. We would like to find the expected
number of pairs of vertices $(P_1,P_2)$ of this cell which do not
share an edge, i.e., which are not consecutive. Any cell vertex $P_i$
is equidistant from three seeds in the sprinkling, in this case $S_0$
and two others, $(S_{i1},S_{i2})$, where $i = 1, 2$ labels the vertex
they are associated with. Therefore, when we count pairs of vertices
we need to count pairs of configurations $(S_0,S_{i1},S_{i2})$ such
that the disk $V_i$ inside the circle through each triple of points
is void of other seeds (so that they really define a vertex), and
the four seeds $S_{ij}$ are distinct (so that the vertices $P_1$
and $P_2$ are not consecutive). What is the probability density
for all of this to happen, in terms of all possible locations for
the four seeds in question?

The probability measure for a seed to be located at $S = (\chi,
\theta)$ is $\rho\,(a^2\,\sin\chi\, \dd\chi\, \dd\theta) = \rho a^2\,
\dd(\cos\chi)\,\dd\theta$; therefore, for the pair of seeds $(S_{i1},
S_{i2})$ giving the vertex $P_i$, at locations $(\chi_{i1},
\theta_{i1})$ and $(\chi_{i2}, \theta_{i2})$, respectively, with
$i = 1$ or 2,
\beq
    \dd\mu_i = \rho^2 a^4\,\dd(\cos\chi_{i1})\,\dd\theta_{i1}
    \,\dd(\cos\chi_{i2})\,\dd\theta_{i2}\;. \label{measure}
\eeq
We can make sure that none of the four seeds is contained in the
disk defined by $S_0$ and the two seeds in the other pair by
specifying appropriate ranges for the allowed positions of the
seeds, and we can impose that the union of the two disks $V_i$ contain
no additional seeds by multiplying the measure by the probability
$\ee^{-\rho A(V_1\cup V_2)}$.

To control more easily the ranges of integration, it is convenient
to make a coordinate transformation from the eight variables
$(\chi_{ij}, \theta_{ij})$ to a set of four variables $(\zeta,
\phi, \omega_1, \omega_2)$ which specify the location of $P_1$
and $P_2$ (or equivalently, the size and location of the two
circles), and four variables $\alpha_{ij}$ which specify the
location of the four points on the two circles.
Given two circles through $S_0$, or two points $P_i = (\chi_i,
\theta_i)$, labeled so that $\theta_2-\theta_1 \le \pi$, call
$Q = (2\zeta, \phi)$ the other point at which the circles intersect,
besides $S_0$ (if the circles are tangent at $S_0$ we identify
$\phi$ with the direction of that tangent, and $Q = S_0$, but
this happens with probability zero). Also, call $\omega_1$ and
$\omega_2$ the angles $\widehat{QS_0P_1}$ and $\widehat{QS_0
P_2}$, taken to be positive respectively in the clockwise and
counterclockwise directions from $\phi$. Then
\beq
     \theta_1 = \phi-\omega_1\;,\qquad \theta_2 = \phi+\omega_2\;,
\eeq
and the ranges of values for the new angles are
\beq
     0 < \phi < 2\pi\;,\quad -{\pi\over2} < \omega_1 < {\pi\over2}\;,
     \quad -\omega_1 < \omega_2 < {\pi\over2}\;.
\eeq
The distance $\chi_i$ of $P_i$ from $S_0$ can be expressed in
terms of $\zeta$ using the spherical cosine law applied to the
isosceles triangle $S_0QP_i$, i.e., $\cos\chi_i = \cos\chi_i
\cos 2\zeta + \sin\chi_i \sin 2\zeta \cos\omega_i$, which
implies the relationship
\beq
     \tan\chi_i = {\tan\zeta \over \cos\omega_i}\;,\qquad
     {\rm with} \qquad 0 < \zeta < {\pi\over2}\;. \label{chi_i}
\eeq

\begin{figure}
  \includegraphics[angle=0,scale=.65]{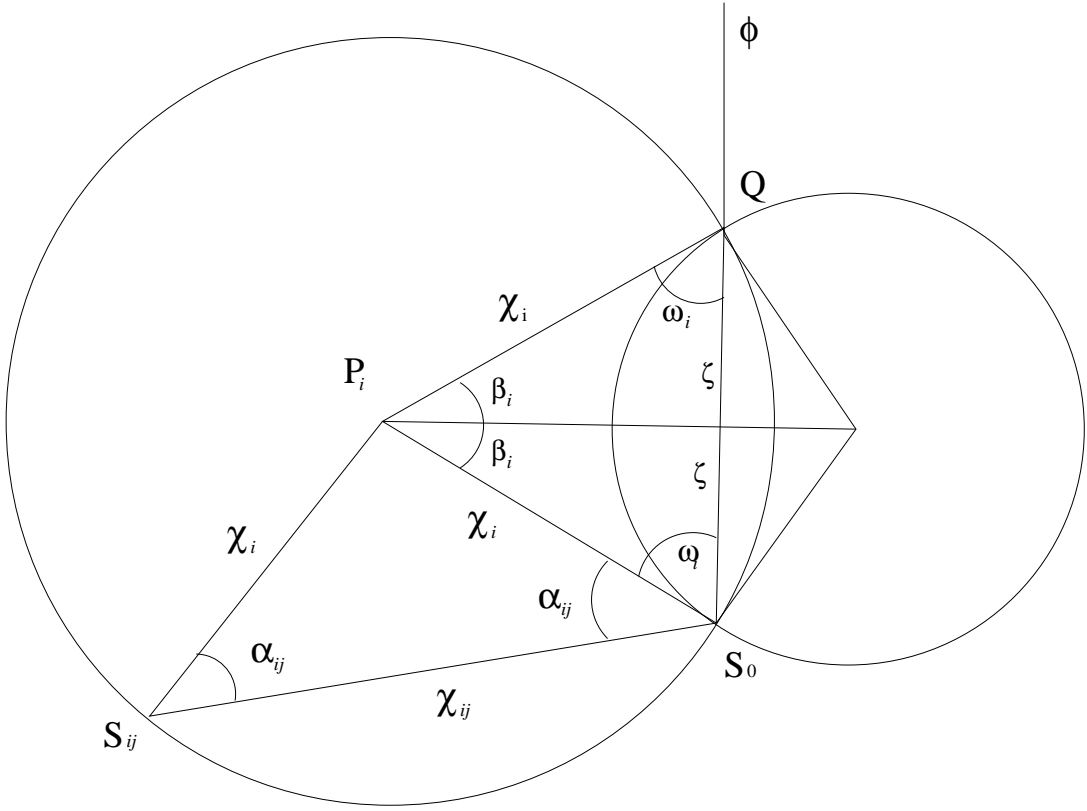}
  \caption{\label{f2}
  Construction for the calculation of 2D statistical
  fluctuations on the two sphere; see text for explanation
  (only one of the two $P_i$ and one of the $S_{ij}$ are labelled).}
\end{figure}

The additional variables are defined as follows. On the circle
around $P_1$ (resp, $P_2$) call $S_{i1}$ the first seed one meets
going around the circle clockwise (resp, counterclockwise) from
$S_0$, and $S_{i2}$ the second one. Then $\alpha_{i1}$ is the
angle $\widehat{P_i S_0 S_{i1}}$ and $\alpha_{i2}$ the angle
$\widehat{P_i S_0 S_{i2}}$ (both measured clockwise for $i = 1$
and counterclockwise for $i = 2$), i.e.,
\beq
     \theta_{ij} = \theta_i + (-1)^i \alpha_{ij}
     = \phi + (-1)^i\,(\omega_i + \alpha_{ij})\;, \label{theta}
\eeq
with ranges given by
\beq
     -\omega_i < \alpha_{i1} < {\pi\over2}\;,\qquad
     \alpha_{i1} < \alpha_{i2} < {\pi\over2}\;.
\eeq
The Eqs (\ref{theta}) constitute half of the coordinate transformation.
To find the other half, we apply the spherical cosine law to the
triangle $S_0P_iS_{ij}$; this gives $\cos\chi_i = \cos\chi_i
\cos\chi_{ij} + \sin\chi_i \sin\chi_{ij} \cos\alpha_{ij}$, which
implies the relationship
\beq
     \tan{\chi_{ij}\over2} = \tan\chi_i \cos\alpha_{ij}\;.
\eeq
We can then substitute (\ref{chi_i}) in this last equation, and
express the result in the more convenient form
\beq
     \cos\chi_{ij} = {\cos^2\omega_i-\tan^2\zeta\, \cos^2\alpha_{ij}
     \over \cos^2\omega_i+\tan^2\zeta\, \cos^2\alpha_{ij}}\;.
     \label{chi2}
\eeq
With (\ref{theta}) and (\ref{chi2}) we can now rewrite the full
measure of integration (\ref{measure}) in terms of the new variables.
After a somewhat lengthy calculation, one obtains
\bea
     \dd\mu_1\,\dd\mu_2
     &=& \dd\zeta\, \dd\phi\, \dd\omega_1\, \dd\omega_2\, \dd\alpha_{11}\,
     \dd\alpha_{12}\, \dd\alpha_{21}\, \dd\alpha_{22}\nonumber\\
     & &\times\ 256\,\rho^4a^8\,\tan^7\zeta\,(1+\tan^2\zeta)
     \left[\prod_{i,j=1,2}{\cos\alpha_{ij}\over(\cos^2\omega_i
     +\tan^2\zeta\, \cos^2\alpha_{ij})^2}\right] \times \nonumber\\
     & &\times\ \cos^3\omega_1\, \cos^3\omega_2\, \sin(\omega_1+\omega_2)\,
     \sin(\alpha_{22}-\alpha_{21})\, \sin(\alpha_{12}-\alpha_{11})\;.
\eea

We now need to calculate the area $A(V_1 \cup V_2)$ of the union
of the disks around $P_1$ and $P_2$. The line $S_0Q$ divides each
of the $V_i$ into two parts, and $V_1 \cup V_2$ is the disjoint
union of one part from each $V_i$ (see figure) which, for the
purpose of finding its area, it is convenient to think of as $V_i$
with the ``wedge" $S_0P_iQ$ removed and replaced by the triangle
$S_0P_iQ$. Thus we can write
\beq
     A(V_1 \cup V_2) = \sum\nolimits_{i=1}^2
     (A_{\rm disk} - A_{\rm wedge} + A_{\rm triangle})_i =
     \sum\nolimits_{i=1}^2 [\pi+2\omega-(\pi-\beta)\cos\chi]\,a^2\;,
\eeq
where some simple spherical geometry gives
\bea
     & &A_{{\rm disk}\;i} = 2\pi\,(1-\cos\chi_i)\,a^2 \nonumber\\
     & &A_{{\rm wedge}\;i} = 2\beta_i\,(1-\cos\chi_i)\,a^2 \nonumber\\
     & &A_{{\rm triangle}\;i} = 2\,(\beta_i + \omega_i - \half\,\pi)\,a^2\;.
\eea
Here, $\beta_i$ is half of the internal angle of the wedge or the
triangle $S_0P_iQ$ at $P_i$. Using the sine law with half the triangle
$QS_0P_i$, we obtain that it is related to our variables by
\beq
     \cos\beta_i = \cos\zeta\,\sin\omega_i\;,
\eeq
and we obtain from (\ref{chi_i}) that
\beq
     \cos^2\chi_i = {\cos^2\omega_i\over\cos^2\omega_i+\tan^2\zeta}\;.
\eeq
Putting these pieces together we therefore have
\beq
     A(V_1 \cup V_2) = a^2 \sum_{i=1}^2 \left[ \pi + 2\,\omega_i
     -{2\cos\omega_i\over\sqrt{\cos^2\omega_i+\tan^2\zeta}} \left(
     \pi - \arccos(\cos\zeta\,\sin\omega_i)\right) \right],
\eeq
and the expectation value we are looking for is obtained integrating
the product of these probabilities over all locations of the four seeds,
\beq
     \ovr{N_{0,0'}} = \int\dd\mu_1\int\dd\mu_2\,
     \ee^{-\rho\, A(V_1\cup V_2)}\;, \label{N00}
\eeq

Finally, substituting this in the expression for the variance of $N_0$, we get
\beq
     \sigma_1^2 = 2\,\ovr{N_{0,0'}} + 3\,\ovr{N_1} - \ovr{N_1}^2
     = 2 \int\dd\mu_1\int\dd\mu_2\, \ee^{-\rho\, A(V_1\cup V_2)}
     -18\left(1 - {3\over4\pi}\,{R\over\rho} + {1\over8\pi^2}
     {R^2\over\rho^2} \right).
\eeq
This integral can be evaluated numerically for any given value of $\rho/R$.

We are now in a position to discuss the geometries we associate
with a 2-dimensional Voronoi complex $\Omega$. Since in two
dimensions the curvature is completely characterized by the Ricci
scalar $R$, our goal is to associate a value of $R$, with suitable
uncertainties, with every set $U$ in a cover of a manifold $\man
\simeq \Omega$ that is made of sufficiently small sets. We start
by selecting a larger collection of candidate sets, and then
explain how to pick the appropriate ones among them. For each
2-cell $\omega_0 \in \Omega$ consider the family of sets
$\{U_{\omega_0,\lambda}\}$, $\lambda = 0$, 1, 2, ..., defined
by
\beq
    U_{\omega_0,0} = \omega_0,\qquad U_{\omega_0,\lambda+1}
    = U_{\omega_0,\lambda} \cup \{\omega \in \Omega \mid \partial\omega
    \cap \partial U_{\omega_0,\lambda} \ne \emptyset\}\;;
\eeq in other words, $U_{\omega_0,\lambda}$ is $\omega_0$ together
with the first $\lambda$ layers of neighboring 2-cells around it.
Roughly speaking, if we call $N_{\omega_0, \lambda}$ the number of
2-cells in $U_{\omega_0, \lambda}$, in an approximately flat
2-geometry we expect to have $N_{\omega_0,\lambda} \approx 1 + 6 +
12 + ... + 6\lambda = 3\lambda^2 + 3\lambda + 1$.
 A value of $R$ can be assigned to each $\omega_0$ simply by using (\ref{Rest}) as
an estimate, \beq
    R(\omega_0) = 4\pi\rho\left(1-\textstyle{1\over6}\,N_1(\omega_0)\right).
\eeq This value however is not a good one as far as manifold
geometry goes; differences between it and those obtained for
neighboring 2-cells should be interpreted not as real variations
of $R$ but as statistical fluctuations. What we should do instead
is average the values obtained for a cluster of neighboring
2-cells, i.e., a suitably large $U_{\omega_0,\lambda}$. If we knew
that the continuum geometry will turn out to be a constant
curvature one, the best strategy would be to pick $\lambda$ as
large as possible, in order to maximize the statistics. In
practice, if we use too many 2-cells we may be combining regions
that in a good fit of the geometry would have different
curvatures, so we need to specify a procedure for picking an
optimal $\lambda$. Pick some value of $\lambda$. Then, at scale
$\lambda$, $U_{\omega_0, \lambda}$ provides a sample of size
$N_{\omega_0, \lambda}$ from the ensemble of all 2-cells in a
manifold of assumed constant curvature, and our best estimate for
the scalar curvature around $\omega_0$ is the average \beq
    R(U_{\omega_0,\lambda})
    = \ev{R(\omega)}_{\omega\in U_{\omega_0,\lambda}}
    = 4\pi\rho\left(1-\textstyle{1\over6}\,
    \ev{N_1(\omega)}_{\omega\in U_{\omega_0,\lambda}}\right),
\eeq with the variance in the distribution of averages over such
samples being given by
\beq
    \sigma_{\ev{N_1}}^2 = N_{\omega_0,\lambda}^{-1}\,\sigma_1^2\;.
\eeq If this variance is small, we can use it to estimate a range
of values of $R$ which would have produced values of
$\ev{N_1(\omega)}$ within this tolerance, by setting $\Delta R =
(\dd R/\dd\ev{N_1})\,\sigma_{\ev{N_1}}$, or
\beq
    \Delta R(U_{\omega_0,\lambda}) = {2\pi\rho\over3}\,\sigma_{\ev{N_1}}
    = {2\pi\rho\over3}\,N_{\omega_0,\lambda}^{-1/2}\,\sigma_1\;.
\eeq

Finally, the whole process is self-consistent and gives a good
continuum approximation to $\Omega$ if differences between the
scalar curvatures estimated from neighboring regions $U_{\omega_0,
\lambda}$ are smaller than the statistical uncertainty $\Delta
R(U_{\omega_0,\lambda})$ within individual regions, i.e., $R$
is slowly varying on the scales given by $\lambda$.

\subsection{Considerations on the General Case}
\label{sec:3.2}

\noindent In three dimensions, there are three independent local
curvature invariants. Different choices are possible; for example,
they can be thought of as the eigenvalues of the Ricci
tensor $R_{ab}$, or as the scalars $R \equiv R^a{}\!_a$,
$R^a{}\!_b\, R^b{}\!_a$, and $R^a{}\!_b\, R^b{}\!_c\, R^c{}\!_a$.
To recover them from $\Omega$ we need to use three independent local
combinatorial graph quantities, chosen among the $N_k(\omega)$.
One may think that a possible choice for these is the set of
numbers $(N_0(C), N_1(C), N_2(C))$, which count the number of
vertices, edges and 2-faces of a 3-cell $C$. However, for any
given $C$, only one of those three numbers is independent, since
its vertices, edges and faces form a 2-dimensional complex
$\partial C$ homeomorphic to S$^2$, to which we can apply Eqs
\ref{euler}--\ref{edges} with $\chi(\partial C) = 2$, and we get
that $N_0(C)-N_1(C)+N_2(C) = 2$ and $N_1(C) = {3\over2}\,N_0(C)$,
which is easily solved to give, for example,
\beq
     N_0(C) = 2\,N_2(C)-4\;,\qquad N_1(C) = 3\,N_2(C)-6\;.
\eeq
It seems reasonable to assume that, analogously to the 2D result,
the number $N_2(C)$ of 2-faces of a 3-cell allows us to estimate
the scalar curvature $R$. The other curvature invariants are
associated with the anisotropies in the curvature, since
knowledge of the traces of powers of $R^a{}_b$ can be thought of
as corresponding to knowledge of its eigenvalues.  We can then
conjecture that this type of curvature information is encoded
in direction-dependent combinatorial quantities of the 3-cells,
i.e., any statistically significant differences between properties
of 2-cells on different parts of the 3-cell boundaries; for
example, the direction in which faces of a 3-cell $C$ have the
least number of edges are those in which $\partial C$ is ``most
curved", and should correspond to an eigenvector of $R^a{}_b$.
Checking such statements is beyond the scope of this article,
and will most likely require computational work with families of
simple 3-geometries.

\section{Small-Scale Geometry and Coarse-Graining}
\label{sec:4}

\noindent In the previous section we assumed that all
complexes are embeddable in some manifold, but one of the
ultimate goals of graph-based approaches to quantum gravity is to
obtain a manifold-independent formulation. This section represents
a tentative first step in that direction. It is somewhat more
speculative than the previous ones, both for this general reason
and for a more specific one. Even though our motivation comes from
the semiclassical sector of quantum gravity, we will continue to work
in a purely geometrical setting, i.e., we do not consider any quantum
aspects, states, operators and the like. What we shall assume is that
it is meaningful to consider a fixed discrete structures even at scales
below the cross-over scale we have considered before, either because
the quantum-to-classical transition occurs at a still smaller scale
$\lq < \ld$, or because we can anyway learn something about the
continuum interpretation of the theory by pushing this geometrical
view as far as possible. This is the meaning we shall give
to the term ``small scale geometry". We will discuss two aspects
of this issue here, the different scales (``combinatorial" scales
$\ell$ in the discrete structure, or metric scales $L$ in the
continuum) that play a role in the construction of a Riemannian
manifold, and how to deal with some of the obstructions that
may come up in this process. As we will see, even in the
simple setting in which we consider only the milder obstructions,
different situations may arise; we will need to introduce the
concept of coarse-graining, and to define other length scales
in addition to the ones we have already seen in other sections.

\subsection{Length Scales}
\label{sec:4.1}

\noindent One general issue that needs to be addressed is that of identifying
the various length scales that appear in our framework, and assigning
dimensional values to combinatorial quantities in $\Omega$. Since
we are assuming that the fundamental theory is formulated in terms
of $\Omega$, we would want to describe those length scales using
quantities intrinsic to $\Omega$, and be able to make statements
such as ``at scale $\lq$, $\Omega$ cannot be embedded in a manifold,"
which implies that a continuum geometry is not available to provide
a meaning for length scales.

Continuum theories include a number of length scales associated
with various phenomena; results of measurements are interpreted in
terms of those scales and provide values for them, so in order to
provide values for scales intrinsic to $\Omega$ it is natural to
go through a correspondence with continuum length scales. An
obvious one to try to identify in terms of $\Omega$ is the Planck
length $\LP$ (when $\Omega$ is dressed with other variables it may
be possible to do this even at the kinematical level, as in the
case of the calculation of the loop spacing for the heuristic
weave states in loop quantum gravity \cite{weaves}), but any
prediction of the discrete theory that has a continuum counterpart
can be used, and when more than one will be available the discrete
theory can be tested.

A discrete structure of the type we are discussing also provides
various scales. For example, if we start with a Voronoi cell complex,
we can embed it in a geometry that has no length scales smaller than
the cell size. The latter cannot be assigned a dimensional value yet,
because the situation is invariant under a global rescaling of the
metric, but we will say that $\Omega$ is associated with a
semiclassical cross-over scale $\ls$ equal to its discreteness
scale $\ld$. If we start with a $D$-dimensional Voronoi complex
$\Omega$ that is not a cell complex, but is such that we can
obtain a cell complex $\Omega'$ by coarse-graining $\Omega$ and
removing a fraction $0 < \xi < 1$ of its $D$-dimensional tiles, we
will say that $\Omega'$ corresponds to the semiclassical scale $\ls$,
at which it and its dual $\Omega'{}^*$ represent the manifold well,
while $\Omega$ is characterized by the discreteness scale $\ld
= (1-\xi)^{1/D}\ls$. In either case, it may not be possible to
determine the metric or curvature of the geometry at scale $\ls$,
because at that scale we do not have enough statistics to reliably
use the techniques we will describe. Instead, the classical scale
$\lc$ is the linear size, in terms of number of cells, of the
regions in $\Omega$ used as approximations to the manifold regions
$X$ such that the continuum geometry can be determined with small
statistical fluctuations. That geometry itself, possibly together
with the global topology of $\man$, may also determine larger
length scales $\lg$ (both local ones defined by the curvature
and its rate of change or higher derivatives, and possibly global
ones defined for example by non-trivial homotopy generators).

Summarizing, the length scales $\ld \le \ls \le \lc \le \lg$ are
determined by the structure of $\Omega$, if $\Omega$ is a discrete
structure of a type that allows us to define $\ls$, given a suitable
definition of a coarse-graining procedure. In quantum gravity, we
will then require, as part of the definition of a semiclassical state,
the condition that $\Psi$ be associated with a discrete structure
for which $\ls$ exists. On the other hand, continuum-based physics
determines a length scale $\LP$ for quantum gravity, given in terms
of fundamental physical constants by dimensional arguments and
back-of-the-envelope calculations of quantum gravity fluctuations
\cite{Wh}, which appears also in the eigenvalues of the quantum
geometry operators for areas and volumes; It is not clear to us
whether this scale should be identified with $\ld$ or $\lc$, if
different, or how this relationship might be affected by
renormalization arguments. Finally, physics also determines a
phenomenological, ``macroscopic" scale $\LM \gg \LP$, which depends
on the experimental techniques used but that can be assumed to be
larger than $\ls$. A good understanding of the relationships between
all of these scales is outside the scope of this paper, but our
discussion of coarse-graining will be a start in this direction.

\subsection{Coarse-Graining}
\label{sec:4.2}

\noindent Some complexes $\Omega$ can be directly associated
with a ``macroscopic" geometry, in the sense that they can be
embedded homeomorphically in a manifold at a length scale $\lc$
and the procedures previously described give a well-defined
(approximate) geometry at a length scale $\ls$, in which case the
``semiclassical" interpretation is relatively straightforward,
within the appropriate statistical uncertainties. That is,
$\Omega$ gives us a classical, mesoscopic geometry. At larger
scales, for which the number of points to be taken per sample cell
is larger, one expects a very good approximation, according to our
own set of conditions. The remaining question then is:
What if $\Omega$ does not satisfy the requirements for
approximating a nice geometry? Although there then is no direct
association between graph and geometry, if in an appropriate sense
the obstructions only occur on small scales, it may be possible to
coarse-grain the graph to a larger-scale one that is associated
with a continuum geometry.

The coarse-graining procedure will be one which, intuitively,
takes a graph $\Omega$ at a certain scale and maps it
to a different, larger-scale $\Omega'$. There are two
possibilities on how to perform the coarse graining in this case.
The first possibility, that we call {\it soft\/} coarse-graining,
has the following strategy: One assumes that one has a Voronoi
or Delaunay complex and one gets to a coarser one by ``ignoring"
the structure below a certain scale simply by averaging the
quantities of interest over neighboring cells. This is what we
have already done in Sec~\ref{sec:3} when looking for the optimal
mesoscopic scale at which to define slowly varying quantities;
we are not actually defining a new complex, and it can always
be done. In the second possibility, that we will refer to as
{\it hard\/} coarse-graining, we have a procedure that connects
two graphs, and one gets the larger graph $\Omega'$ by removing
cells (Delaunay vertices) from $\Omega$, in a precise way.

Hard coarse graining, which implies a modification of the
underlying graph, represents a new input in our considerations.
This procedure is achieved by a sequence of {\it cellular
moves\/} that refine the cellular decomposition \cite{moves}.
These moves have been shown to provide a prescription for
refining and coarsening the complex (they are in a sense
more elementary than the Pachner moves). The intuitive idea
for the coarse graining is that two adjacent $n$-cells of the
complex that share an $(n-1)$-cell get `fused' into a new
$n$-cell. This refining/coarsening procedure has already been
used for defining renormalization prescriptions in discrete
systems \cite{renorm}, and the approximation to smooth
manifolds by our proposed hard coarse-graining is an explicit
example of some of those prescriptions.

In order to have a proper understanding of the different scenarios
that one might encounter given an arbitrary graph at small scales,
with the purpose of incorporating it into our geometrical
formalism, let us make the following classification in decreasing
order of `complexity':

\begin{enumerate}

\item In the most exotic scenario one starts with a Voronoi
complex that does not have a dual cell complex. This means that
there is no dual triangulation and the corresponding graph might
be disconnected. Since this happens for cases where, say, one
sprinkles points in a manifold with a density that is ``too low",
and no real statistics is possible, we will exclude this case
from our considerations here.

\item The next possibility is that the given cell complex $\Omega$
can not be embedded into any manifold $\man$. This means that the
complex is not topologically a cell decomposition of a manifold.
In this case one might hope that with a proper coarse graining one
might be able to take this complex to a new one $\Omega'$ that
does admit such interpretation and can be embedded. Physically,
one could say that the original complex represented a space that
on certain small scales has `topological' and/or `dimensional'
fluctuations. This is the trickiest situation where one needs a
`hard coarse graining', for which we are not aware of an existing
specific procedure, and it is certainly a question worth pursuing.

\item A milder scenario is to consider as the starting point a
cell complex $\Omega$ that is a cell decomposition of a manifold
$\man$ but that is not Voronoi, in the sense that the valence of
the graph is not equal to $D+1$. In this scenario, one would allow
for ``fluctuating valence'' in the Voronoi cell, but with average
$\ev{N_{1|0}} = 4$, in 3D. This means that the graph that defines
the quantum state at Planck scale does not have the `Voronoi
signature', which might be the case for an arbitrary graph in loop
quantum gravity. In this case we can implement the cell moves
of Ref \cite{moves} to turn it into a Voronoi cell complex; if
the authors' conjecture is correct, via these cellular moves one
can always transform any cell decomposition into a Voronoi one,
that will serve as starting point for the statistical geometry
considerations described in previous sections. One should also
note that in some cases this `restructuring' of the cell complex
might not correspond to a course graining in the strict sense,
since the moves that `lower' the valence of the Voronoi graph
do not remove cell complexes to it as one might expect from a
coarse graining procedure.

\end{enumerate}

In the next, final section we shall consider quantum issues
and relate the formalisms we have been employing to the task
of bridging the gap between a semiclassical state in loop
quantum gravity and classical geometry.

\medskip

\section{Planck Scale Geometry and Beyond}
\label{sec:5}

\noindent In this section we make precise the connection between
the previous sections and semiclassical canonical quantum gravity,
i.e., the low energy limit of loop quantum gravity. This section
has two parts. In the first one, we outline how one can try to use
our framework to complement present approaches to semi-classical
states in LQG. In the second part we outline the steps to be
followed in order to estimate statistical errors in making
phenomenological predictions out of the semiclassical states.

\subsection{Semiclassical States}

\noindent In order to deal with the low energy limit of the theory,
we will again have to deal with the fact that there might be new
length scales in the description of the quantum geometry. So far we
have considered in great detail the way in which we may approximate
a macroscopic geometry starting from a graph or complex at a much
smaller scale. Even though we have introduced a microscopic
discrete scale $\ld$, at which the finest graph might be defined,
this scale is not necessarily related to a particular quantum
scale $\lq$. As discussed in Sec~\ref{sec:2}, there are standard
dimensional arguments which indicate that the relevant scale $\lq$
for a quantum state is the Planck length $\LP$. This only means
that the quantum excitations of the geometry, which could be
elements of area through edges of the graph or contributions to
the volume from the vertices, yield eigenvalues of the order of
the Planck scale for single edges/vertices (at least for the
unrenormalized value of the Barbero-Immirzi parameter $\gamma$ of
order one that is used nowadays). Whenever we consider a
semiclassical state $\Psi_{\rm sc}$ in loop quantum gravity, there
are several steps and assumptions involved in estimating the
continuum geometry it corresponds to. Let us now explore what
those assumptions are.

The first assumption is that the true nature of geometry at the
Planck length is described by loop quantum geometry. Second, we
assume that we can define certain states (Coherent
\cite{toherent}, Shadow \cite{shadow}, Gaussian \cite{gaussian},
etc.) that will be semiclassical in the sense of approximating a
geometry (and its time derivative), in the `correct phase'.
Finally, one is assuming that a definite picture of the quantum
geometry will emerge from a precise merger of both loop quantum
gravity and our ``discrete-statistical'' approach. In this
viewpoint, the final picture looks like this: The `true state' is
given by a properly defined state in loop quantum gravity,
featuring quantum behavior at the Planck scale. The graph that we
use to define this `shadow state' (to give an example) will be a
Voronoi graph. The semiclassical state will have the property
that, when probed on a mesoscopic scale $\ell_{\rm c}$ (still to
be specified in practice) it will behave quasi-classically
(hopefully, with respect the the coarse-grained operators that we
need to define). This means that in order to construct the desired
semiclassical state following the steps of Ref~\cite{ABC}, the
operators that we specify as belonging to the set to be
approximated (together with the tolerances) will {\it not\/} be
operators assigned to observables at the quantum, Planck scale.

We are interested in constructing states that approximate
quantities at a scale $\ell_{\rm s}$, where the quantum-classical
transition takes place. As already mentioned, this new scale, yet
to be identified, could be assumed to be close to the mesoscopic
scale $\ell_{\rm c}$, but it should be smaller. Going to our analogy
model of fluids, the transition quantum-classical can be assumed
to take place at a smaller scale than the one at which averages
are taken and the continuum approximation emerges (for the simple
reason that one adds velocities of particles contained in the
region, which already presupposes classical attributes for the
constituents of the fluid). The viewpoint that new, mesoscopic
scales have to be considered has already been explored by other
authors who have explored the semiclassical limit before
\cite{semiclassical}.

The next question we would like to address is the possible utility
of our statistical methods for distinguishing between good and bad
candidates for semiclassical states. That is, suppose that we are
given an alleged semiclassical state $\Psi_{\rm sc}$ and we are
assigned the task of testing it. The strategy for doing so is the
following. Compute expectation values and fluctuations for the
cell-based,  observables that the maker of the state claims it
approximates well. Make sure that the state is defined on a
Voronoi complex. Dress the complex with the expectation values
found for the observables. Now the crucial step is to compare this
dressing with the geometry we expect to get from the complex
itself, as in Sec~\ref{sec:3}, and verify that they are
consistent. If they are not, return the state to the maker.

\subsection{Phenomenology}

\noindent Suppose now that the given state satisfies all criteria set
by some considerations, then in our procedure we would have to pass
it to the next step in the testing line: see whether it fits
observations. Now this is a very intricate question, for we have
not specified the phenomenological criteria that a semiclassical
state should satisfy, nor the expected observable deviations from
the classical realm (Lorentz violation, modified dispersion
relations, non-commutativity, birefringence, etc)
\cite{phenomeno}. Note also that this procedure implies the
knowledge of more quantities assigned to the cell complex, more
than what we have up to now specified in the complex. This further
`dressing of the complex' will be dealt with in a forthcoming
paper \cite{bcw2}.

What we could attempt to answer is the following simpler question.
Suppose that we expect the state to approximate the geometry
$g_{ab}$. As we have discussed before, there are now two sources
of uncertainty in the approximation of this spacetime geometry
(even though we were working in the canonical picture, we can talk
about a spacetime geometry). The first source is a well known one,
due to quantum fluctuations: observables $O_I$ that were chosen to
be approximated by the state have quantum dispersions $(\Delta
\hat O_I)_\Psi$. The second source of uncertainty is the
statistical nature of our reconstruction procedure. That is, as we
have argued before, the macroscopic continuous geometry is only a
fiction, very useful for describing phenomena at certain scales,
but it is only an approximation to the true geometry. But there is
not a unique macroscopic metric that can be approximated by the
graph: rather, there is an `ensemble' of such metrics, which then
introduces a classical or cross-over uncertainty $(\Delta
O_I)_{\rm c}$ in the observables. The observational imprint that
this uncertainty might bring is manifested by the fact that we
normally use a fiducial classical metric to perform geometrical
measurements. We expect that this geometry belongs to the set of
possible geometries that are approximated by the graph, but it
might turn out to be not the `most probable' one, and thus one
might be introducing an extra error in our interpretation of the
experimental results. Note that these errors are of a different
nature than those that arise from the use of a regular lattice in
\cite{TH}. We expect that we won't have these `discretization
errors' due to the use of the Voronoi construction. In order to
have full control over the different sources of uncertainty, one
would like to quantify them and decide which one is the dominant
one. At this point we are not in the position of having a working
hypothesis for this question, and we can only make an educated
guess as to the way in which the two uncertainties interact with
each other. Under the most naive assumption of independence, one
might say that the total uncertainty is
$$
   (\Delta O_I)^2_{\rm t} = (\Delta
   O_I)^2_\Psi + (\Delta O_I)^2_{\rm c}\;.
$$
An important issue is to quantify the fluctuations that can be
attributed to both effects, and know how the total fluctuation
for each observable depends on them, since we would like to be
able to subtract any statistical `spurious' information and be
able to measure the pure quantum contributions to the problem.

Assuming that the semiclassical state had these desired properties
---an issue that lies outside the scope of this paper--- would lead
us to conclude, from a strict viewpoint that we have succeeded in
our task of constructing a semiclassical geometry. This is
because this state should then approximate any observations, at
the mesoscopic level that we shall perform on the geometry (with
external matter) and thus, if we now take the viewpoint that it is
enough to describe what we can observe, we would reach our
conclusion.

\subsection{Summary and Outlook}

\noindent To summarize, we have introduced a framework for establishing
a bridge between the description of geometry of space in terms of
semiclassical states for quantum gravity, and the one used in
phenomenological calculations of corrections to field theory in
curved space. At the most basic level, our proposal fits in with
the general idea that the type of discreteness one encounters in a
class of approaches to quantum gravity implies that the amount of
information contained in a finite spacetime volume is finite,
which has been explored in a variety of contexts \cite{info}. Our
work then suggests a specific way to translate this idea into
quantitative expressions for the information a discrete structure
contains about a continuum geometry, using statistical geometry.
To illustrate how the framework is used, we have characterized a
semiclassical quantum gravity state by a single cell complex and a
single set of values for variables motivated by loop quantum
gravity. Such a characterization, of course, is not intended to be
complete; it might arise as part of the specification of a
coherent state \cite{toherent}, or from a ``shadow state" used to
probe $\Psi_{\rm sc}$ \cite{shadow}, but for our purposes it is
just a tool that allows us to isolate the statistical effects we
want to study. It is a tool that complements the criteria one
might have in choosing the quantum state.

The general picture that emerged in this framework is one in
which, from a certain length scale upward, if the quantities
characterizing the state are sufficiently regular, the mesoscopic
geometry is determined up to quantifiable uncertainties. Once
those uncertainties will be better understood, especially in three
dimensions, one can begin to look into the possibility of adding
matter coupled to gravity, and how these uncertainties need to be
combined with quantum fluctuations to give measurable effects on
observable quantities. The answer to this last question will
probably depend on which observations are carried out, and it is
in principle possible that one can use observable effects that
depend only on quantum uncertainties; after all, going back to the
gas analogy, when we want to study the quantum properties of
atoms, we do not need to combine their quantum fluctuations with
the statistical mechanics description of a gas. But a gas is
different, and we do not need to have information about the state
of the whole gas to even be able to describe what each atom does.

In this paper we have concentrated our attention to graphs that
did not have any other geometrical information other than what is
already in its definition. In order to have an approach closer to
the variables used in LQG one would like to have assigned certain
(classical) quantities to the graph $\Omega$, that is, one would
like to consider `dressed' graphs. This will be dealt with in a
subsequent publication \cite{bcw2}.

\section*{Acknowledgements}

\noindent LB and AC would like to thank Perimeter Institute for
hospitality. Partial support for this work was provided by NSF
grant PHY-0010061, CONACyT grant J32754-E and DGAPA-UNAM grant
112401.
\raggedright


\begin{thebibliography}{999}

\bibitem{conceptual} C Rovelli 2001
``Quantum space-time: What do we know?'' in C Callender (ed)
{\sl Physics Meets Philosophy at the Planck Scale} 101--122,
{\tt arXiv:gr-qc/9903045};
C Rovelli 2000 ``The century of the incomplete revolution:
Searching for general relativistic quantum field theory,''
{\sl J.\ Math.\ Phys.}\ {\bf41} 3776,
{\tt arXiv:hep-th/9910131}.

\bibitem{Wh}J A Wheeler 1962 {\sl Geometrodynamics\/} Academic Press.

\bibitem{lqg} A Ashtekar and J Lewandowski 2004
``Background independent quantum gravity: A status report,'' {\tt
arXiv:gr-qc/0404018};
C Rovelli ``Quantum Gravity", (Cambridge U. Press, 2004); C
Rovelli 1998 ``Loop quantum gravity,'' {\sl Living Rev.\ Rel.}\
{\bf1} 1 {\tt arXiv:gr-qc/9710008};
T Thiemann 2001 ``Introduction to modern canonical quantum general
relativity,'' {\tt arXiv:gr-qc/0110034}.


\bibitem{regge} R M Williams and P A Tuckey 1992
``Regge calculus: A Bibliography and brief review,''
{\sl Class.\ Quantum Grav.}\ {\bf9} 1409;
T Regge and R M Williams 2000 ``Discrete structures in gravity,''
{\sl J.\ Math.\ Phys.}\ {\bf41} 3964, {\tt arXiv:gr-qc/0012035}.

\bibitem{loll} R Loll 1998
``Discrete approaches to quantum gravity in four dimensions,''
{\sl Living Rev.\ Rel.}\ {\bf1} 13, {\tt arXiv:gr-qc/9805049}.


\bibitem{spinfoam} A P\'erez 2003
``Spin foam models for quantum gravity,'' {\sl Class.\ Quantum Grav.}\
{\bf20} R43, {\tt arXiv:gr-qc/0301113};
D~Oriti 2003 ``Spin foam models of quantum spacetime,''
{\tt arXiv:gr-qc/0311066}.

\bibitem{posets} L Bombelli, J Lee, D Meyer and R D Sorkin 1987 ``Space-time
as a causal set" {\sl Phys.\ Rev.\ Lett.}\ {\bf59} 521--524; R D Sorkin 2003
``Causal sets: Discrete gravity (Notes for the Valdivia Summer School),"
{\tt arXiv:gr-qc/0309009}.

\bibitem{bcw2} L Bombelli, A Corichi and O Winkler 2004 in preparation.

\bibitem{Lee} N H Christ, R Friedberg and T D Lee 1982
``Random lattice field theory: General formulation" {\sl Nucl.\
Phys}. {\bf B202} 89--125; T D Lee 1985 ``Discrete mechanics", in
A Zichichi, ed {\sl How Far Are We from the Gauge Forces\/} (XXI
Ettore Majorana School, Erice 1983) Plenum.

\bibitem{TH}
H Sahlmann and T Thiemann 2002 ``Towards the QFT on curved
spacetime limit of QGR. I: A general scheme,'' {\tt
arXiv:gr-qc/0207030};
H Sahlmann and T Thiemann 2002 ``Towards the QFT on curved
spacetime limit of QGR. II: A concrete implementation,'' {\tt
arXiv:gr-qc/0207031};

\bibitem{vol} A Ashtekar and J Lewandowski 1997
``Quantum theory of geometry. II: Volume operators" {\sl Adv.\
Theor.\ Math.\ Phys.}\ {\bf1} 388--429, {\tt gr-qc/9711031}.

\bibitem{Sa} L Santal\'o 1976 {\sl Integral Geometry and Geometric
Probability\/} Addison Wesley.

\bibitem{ID} C Itzykson \& J M Drouffe 1989 {\sl Statistical Field Theory}
Cambridge University Press.





\bibitem{Bo} L Bombelli 2002 ``Statistical geometry of random weave states"
{\sl Proceedings of the Ninth Marcel Grossmann Meeting on General Relativity}, eds V G Gurzadyan, R T Jantzen and R Ruffini, World Scientific, {\tt arXiv:gr-qc/0101080}.

\bibitem{shadow} A Ashtekar and J Lewandowski, 2001
``Relation between polymer and Fock excitations,'' {\sl Class.\
Quantum Grav.}\ {\bf18} L117, {\tt arXiv:gr-qc/0107043}.

\bibitem{JMP} L Bombelli 2000 ``Statistical Lorentzian geometry and the
closeness of Lorentzian manifolds" {\sl J. Math.\ Phys.}\ {\bf41}
6944--6958, {\tt arXiv:gr-qc/0002053}

\bibitem{Br} K Brakke 1985 ``Statistics of random plane Voronoi
tessellations," preprint.

\bibitem{weaves} A~Ashtekar, C~Rovelli and L~Smolin 1992
``Weaving a classical geometry with quantum threads,'' Phys.\
Rev.\ Lett.\  {\bf 69} 237  {\tt arXiv:hep-th/9203079};
 N~Grot and C~Rovelli 1997
``Weave states in loop quantum gravity,'' Gen.\ Rel.\ Grav.\  {\bf
29} 1039.

\bibitem{moves} F Girelli, R Oeckl and A Perez 2002
``Spin foam diagrammatics and topological invariance'' {\sl
Class.\ Quant.\ Grav.}\  {\bf 19}, 1093 {\tt arXiv:gr-qc/0111022}

\bibitem{renorm}
R Oeckl 2003 ``Renormalization of discrete models without
background,'' {\sl Nucl.\ Phys.}\ {\bf B657} 107, {\tt arXiv:gr-qc/0212047};
R Oeckl 2004 ``Renormalization for spin foam models of quantum
gravity,'' {\tt arXiv:gr-qc/0401087};
F Markopoulou 2003 ``Coarse graining in spin foam models,''
{\sl Class.\ Quantum Grav.}\ {\bf20} 777, {\tt arXiv:gr-qc/0203036}.

\bibitem{ABC} A Ashtekar, L Bombelli and A Corichi 2004 ``Coherent and
semiclassical states for constrained systems," in preparation.

\bibitem{Mi} W A Miller 1997 ``The Hilbert action in Regge calculus"
{\em Class. Quantum Grav.} {\bf14} L199--L204, {\tt
arXiv:gr-qc/9708011}.

\bibitem{Ok} A Okabe et al 1999 {\sl Spatial Tessellations\/} 2nd ed, Wiley.

\bibitem{toherent} H Sahlmann, T Thiemann and O Winkler, 2001
``Coherent states for canonical quantum general relativity and the
infinite tensor product extension,'' {\sl Nucl.\ Phys.}\ B {\bf606}
401. {\tt arXiv:gr-qc/0102038}.

\bibitem{semiclassical}
M Bojowald and H A Morales-T\'ecotl 2003 ``Cosmological applications
of loop quantum gravity,'' {\tt arXiv:gr-qc/0306008};
J \'Alfaro, H A Morales-T\'ecotl and L F Urrutia 2002
``Loop quantum gravity and light propagation,''
{\sl Phys.\ Rev.}\ D {\bf 65}: 103509, {\tt arXiv:hep-th/0108061};
J \'Alfaro and G Palma 2003 ``Loop quantum gravity and ultra high
energy cosmic rays,'' {\sl Phys.\ Rev.}\ D {\bf67}: 083003,
{\tt arXiv:hep-th/0208193};
J \'Alfaro, H A Morales-T\'ecotl and L F Urrutia 2002
``Quantum gravity and spin-$\half$ particles effective dynamics,''
{\sl Phys.\ Rev.}\ D {\bf66}: 124006, {\tt arXiv:hep-th/0208192};

\bibitem{gaussian}
A Corichi and J M Reyes 2001 ``A Gaussian weave for kinematical
loop quantum gravity,'' {\sl Int.\ J.\ Mod.\ Phys.}\ D {\bf10}
325, {\tt arXiv:gr-qc/0006067}.

\bibitem{phenomeno} T Jacobson, S Liberati and D Mattingly 2004
``Quantum Gravity Phenomenology and Lorentz violation,'' {\tt
arXiv:gr-qc/0404067}; G Amelino-Camelia 2004 ``Some encouraging and
some cautionary remarks on doubly special relativity in quantum
gravity,'' {\tt arXiv:gr-qc/0402092};
R C Myers and M Pospelov 2004 ``Experimental challenges for quantum
gravity,'' {\tt arXiv:gr-qc/0402028};
J Collins, A P\'erez, D Sudarsky, L Urrutia and H Vucetich 2004
``Lorentz invariance: An additional fine-tuning problem,''
{\tt arXiv:gr-qc/0403053}.

\bibitem{info} See, e.g., A Kempf 2003 ``A covariant information-density cutoff in curved space-time" {\tt arXiv:gr-qc/0310035}.

\end{thebibliography}
\end{document}